\documentclass{amsart}
\usepackage{graphicx,xcolor,hyperref} 
\DeclareGraphicsExtensions{.png,jpg,.pdf,.eps}

\usepackage{booktabs}

\newcommand{\G}{\mathcal{G}}
\newcommand{\C}{\mathcal{C}}
\newcommand{\LL}{\mathcal{L}}
\newcommand{\tL}{\tilde{\mathcal{L}}}
\newcommand{\tlL}{\tilde{\mathcal{L}}_\text{lin}}

\newcommand{\w}{\mathfrak{w}}
\newcommand{\sss}{\mathfrak{s}}
\newcommand{\n}{\mathfrak{n}}
\newcommand{\PP}{\text{PP}}
\newcommand{\Schw}{\text{Schw}}
\newcommand{\tPP}{\tilde{\text{PP}}}
\newcommand{\tSchw}{\tilde{\text{Schw}}}

\theoremstyle{definition}

\title{Revisiting compactness for district plans}
\author{Kristopher Tapp}
\date{\today}

\begin{document}

\begin{abstract} 
Modern sampling methods create ensembles of district maps that score well on \emph{discrete} compactness scores, whereas the Polsby-Popper and other \emph{shape-based} scores remain highly relevant for building fair maps and litigating unfair ones.  The aim of this paper is twofold.  First, we introduce population-weighted versions of shape-based scores and show a precise sense in which this interpolates between shape-based and discrete scores.  Second, we introduce a modification of the ReCom sampling method that produces ensembles of maps with improved shape-based compactness scores. 
\end{abstract}

\maketitle

\section{Introduction}\label{S:intro}
A variety of elected bodies, including the United States House of Representatives and state legislative chambers,  are formed by partitioning a state or other region into districts and then electing one representative from each district.  Many states have laws that require their congressional and legislative districts to be \emph{compact}, which is typically not defined very precisely but is meant to disallow the convoluted tentacled shapes commonly associated with gerrymandering.  Precise mathematical measurements of compactness are of two types.  First, classical \emph{shape-based} compactness scores are based on perimeter and area measurements.  These include the Polsby-Popper score, the Reock score, and many others.  

Second, \emph{discrete} compactness scores, including the \emph{cut edge count} and the \emph{spanning tree score},  only recently entered the redistricting conversation.  In fact, the term \emph{discrete} means that the measurement depends only on the dual graph, which is defined below and is the starting point of modern computational redistricting algorithms, including the ReCom algorithm introduced in~\cite{ReCom}.  The \emph{ensembles} of thousands or millions of random maps produced by these algorithms have served as baselines against which proposed and enacted maps have been compared in many recent court cases.

Discrete compactness scores have certain advantages over shape-based scores, as discussed in~\cite{Moon_Bridget}.  Moreover, discrete scores cannot be ignored because ReCom and other sampling methods have been empirically and theoretically shown to select maps with good discrete compactness scores \footnote{More precisely, the sequential method in~\cite{SMC} and the reversible Markov chain methods in~\cite{MergeSplit} and~\cite{Rev_ReCom} have been shown to draw maps proportional to their \emph{spanning tree score}, which in turn has been shown to be strongly related to cut edge count; see~\cite{Tapp} and~\cite{Arial_Jamie}.}.

One of the simplest shape-based measurements of the compactness of a map is its \emph{total perimeter}, which means the combined perimeter of all of the district boundary lines, not including the exterior boundary of the state.  This measurement bears a certain similarity to the cut edge count.  In fact, one purpose of this paper is to precisely quantify the differences between the shape-based total perimeter score and the discrete cut edge count, and to construct intermediate scores that interpolate between them.  In particular, if we perform a certain \emph{conformal change} to the underlying metric of the state before computing the total perimeter, we obtain an intermediate measurement that shares the population-weighted feature that the cut edge count naturally possesses. This addresses a criticism from~\cite{Moon_Bridget} of the way that shape-based scores prioritize area rather than population. 

This paper is organized as follows.  In Section 2, we define the dual graph and the relevant compactness scores.  In Section 3, we introduce a conformal (population-weighted) version of the total perimeter score, and exhibit a precise sense in which it interpolates between the total perimeter and the cut edge count.  In Section 4, we borrow an idea from~\cite{Colorado} to introduce variants of the ReCom sampling method designed to generate ensembles of maps with improved scores on various compactness measurements.

All code and data for this paper are available at \url{https://github.com/KrisTapp/Compactness}
\section{Setup}
The typical starting point of computational redistricting is a tiling of a state by atomic units that we will generically call \emph{precincts}.  They might be voting districts (VTDs), census blocks, counties, or other units.  A \emph{map} is a partition of these precincts into contiguous districts that meets state-specific requirements including population balance. 

The \emph{dual graph}, $\G$, of the tiling contains one node for each precinct and an edge between a pair of nodes whenever the corresponding precincts are adjacent (which means they share a boundary of positive length).  A measurement is called \emph{discrete} if it depends only on $\G$.  For example, the size of the set, $\C$, of \emph{cut edges} (edges of $\G$ that connect two nodes located in different districts)  is a natural discrete measurement of the compactness of a given map.   

Since we can write $|\C| = \sum_{e\in\C} 1$, the shape-based measurement best aligned with this discrete measurement is the \emph{total perimeter}, $\LL = \sum_{e\in\C} l(e)$, where $l(e)$ denotes the perimeter of the shared boundary between the precincts connected by $e$.  

\begin{figure}[bht!]\centering
\includegraphics[width=5in]{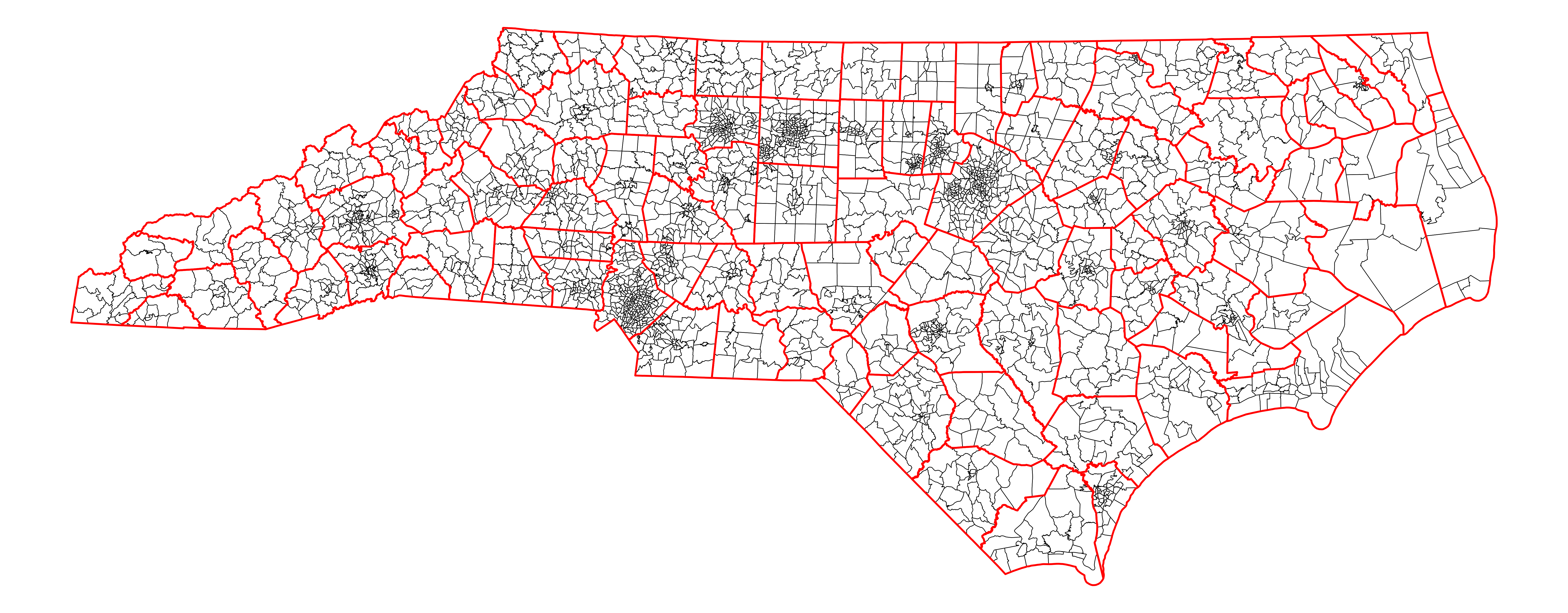}
\caption{The 2692 VTDs of North Carolina with county lines shown in red.}\label{F:NC_precincts}
\end{figure}

One difference between $|\C|$ and $\LL$ is that $|\C|$ tends to be population-density weighted; that is, $|\C|$ imposes a higher penalty on district boundaries through highly populous regions.  This is because such regions tend to have more densely packed precincts, at least in typical applications in which the atomic units are VTDs; for example, Figure~\ref{F:NC_precincts} illustrates the dense packing of VTDs in the urban areas of North Carolina.  The first goal of the next section is to make a variant of $\LL$ that is population-density weighted and is thereby more like $|\C|$.

\section{Total perimeter after a conformal change}
In this section, we apply a conformal change to the underlying metric of the state before computing the total perimeter.  In general, a conformal change by a conformal factor $\phi$ has the effect of multiplying infinitesimal lengths by $\sqrt{\phi}$ and infinitesimal areas by $\phi$.  The conformal factor $\phi$ that we will use is the population density function, so that area in the new metric becomes equal to population.

Although \emph{continuous} conformal factors are standard in Differential Geometry, we only have discrete data available, so we will compute the conformally modified version of $l(e)$ by approximating $\phi$ as a constant on the adjacent pairs of precincts connected by $e$.  More specifically, for each $e\in\C$, we approximate $\phi$ by the following constant value over the entire adjacent pair of precincts connected by $e$:
$$\phi(e) = \frac{P(e)}{A(e)},$$
where $P(e)$ and $A(e)$ respectively denote the combined population and combined area of the precinct pair.  The conformally modified version of $\LL$, which we call the \emph{total conformal perimeter}, is
\begin{equation}\label{E:defL}\tL = \sum_{e\in\C}\sqrt{\phi(e)}\cdot l(e).\end{equation}

One of the main criticisms in~\cite{Moon_Bridget} of shape-based compactness measurements is their emphasis on area rather than population.  The measurement $\tL$ avoids this deficiency by using a conformal change that converts areas into populations.

\begin{figure}[bht!]\centering
\includegraphics[width=1.5in]{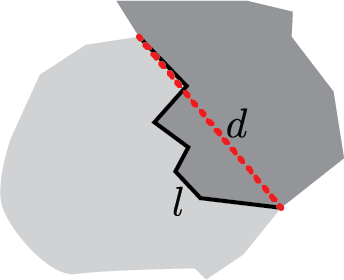}
\caption{The perimeter $l$ and Euclidean distance $d$ of the curve along which two precincts intersect. The \emph{slack} $\sss = \frac{l}{d}\in[1,\infty)$ is a scale-invariant measurement of how wiggly the boundary is.}\label{F:precinct_pair}
\end{figure}

We now wish to quantify the ways in which $\tL$ and $|\C|$ still differ.  The most obvious is that only $\tL$ is sensitive to wiggles in precinct boundaries.  In other words, only $\tL$ is sensitive to the \emph{slack}, $\sss(e) = \frac{l(e)}{d(e)}\in[1,\infty)$, where $d(e)$ denotes the Euclidean distance between the endpoints of the curve along which the two precincts intersect\footnote{If the intersection of the precincts comprises multiple disjoint curves, then $d(e)$ is summed over the components.  In North Carolina, this occurs for only about $0.2\%$ of the dual graph's edges.}, as shown in Figure~\ref{F:precinct_pair}.  Therefore, a natural stepping stone score between $\tL$ and $|\C|$ is the \emph{total conformal linear perimeter} defined as
\begin{equation}\label{E:deflL}\tlL=\sum_{e\in\C}\sqrt{\phi(e)}\cdot d(e) = \sum_{e\in\C}\sqrt{\phi(e)}\cdot\frac{l(e)}{\sss(e)}.\end{equation}

Comparing Equations~\ref{E:defL} and~\ref{E:deflL}, we see that $\tL$ and $\tlL$ differ only in whether they are sensitive to the slack $\sss$. Some disadvantages of slack-sensitivity are discussed in~\cite{Moon_Bridget}. 
 For example, one might not want to penalize a map for following natural boundaries like rivers, which are often good choices for district boundaries.  Furthermore, especially when small atomic units are used, one might prefer to perform measurements with a yardstick that does not detect wiggles at the atomic level of magnification. 

How do the measurements $\tlL$ and $|\C|$ still differ?  To address this, we first consider the degree to which each of $\{\LL, \tL, \tlL, |\C|\}$ depends on the precinct tiling.  First, $\LL$ has no dependence -- it only depends on the boundary lines.  Second, if $\phi$ were modeled as a continuous function, then $\tL$ would also depend only on the boundary lines.  Thus, $\tL$ has only a mild dependence on the tiling caused by the need to discretely approximate $\phi$.

To understand the dependence of $\tlL$ on the precinct tiling, note that $\tlL\leq\tL$ and that $\tlL\rightarrow\tL$ in the limit as the precinct tiling is repeatedly refined.  The shapefile data from the Census Bureau approximates each precinct (and hence each boundary between precinct pairs) as a piecewise linear path. Passing from $\tL$ to $\tlL$ just coarsens this piecewise-linear approximation to the magnification level of precincts.  On the other hand, the cut edge count $|\C|$ depends more significantly on the precinct tiling; in particular, $|\C|\rightarrow\infty$ as the precinct tiling is repeatedly refined.  

In summary, a primary difference between $\tlL$ and $|\C|$, in vague terms, is the degree of dependence on the precinct tiling.  To make this more precise, the following equation expresses $\tlL$ not just as a modification of $\LL = \sum_{e\in\C}\textcolor{red}{l(e)}$ that makes it more like $|\C| = \sum_{e\in\C}\textcolor{red}{1}$, but also as the reverse --  a \emph{weighted cut edge count} whose weighting eliminates some of the dependence on the precinct tiling and hence makes the measurement more like the total perimeter.  
\begin{gather}\label{E:two}
\tlL     =\sum_{e\in\C} \underbrace{\sqrt{\frac{P(e)}{A(e)}}}_{\sqrt{\phi}}\cdot
            \frac{1}{\sss(e)}\cdot \color{red}l(e) \color{black} 
        = \sum_{e\in\C}
        \underbrace{\sqrt{P(e)}}_{\n(e)}\cdot
        \underbrace{\frac{d(e)}{\sqrt{A(e)}}}_{\w(e)=\text{waist}}
        \cdot \color{red}1\color{black}.
\end{gather}

\begin{figure}[bht!]\centering
\includegraphics[width=3in]{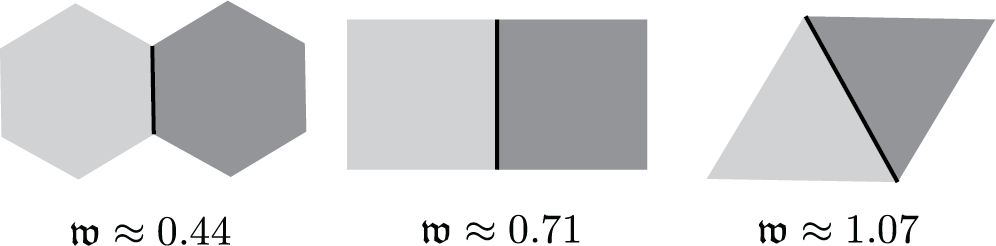}
\caption{The waist $\w$ of three regular lattices.}\label{F:waist}
\end{figure}

Equation~\ref{E:two} parses apart and names the two geometric sensitivities that are baked into $|\C|$ but not into $\tlL$, namely $\n(e)$ and $\w(e)$, whose interpretations are best discussed from the viewpoint that $\tlL = \sum_{e\in\C} \n(e)\cdot\w(e)\cdot\color{red}1\color{black}$ is a weighted version of the cut edge count.
\begin{enumerate}
    \item The measurement $\n(e)=\sqrt{P(e)}$ reflects the precinct population density in the sense that $\frac{\n^2}{2}$ is the local number of people per precinct.  The weighted score $\sum_{e\in\C}\n(e)\cdot 1$ attempts to weight the cut edges so the measurement becomes approximately proportional to the number of cut edges there would be if the precincts were subdivided and/or merged to become equipopulous.  For example, both precincts in a precinct-pair that is $9$ times too populous would be subdivided into a $3$-by-$3$ grid, so each original cut edge would roughly correspond to $3$ cut edges in the new configuration.
    
    \item The ``waist'' $\w(e)=\frac{d(e)}{\sqrt{A(e)}}$ is a scale-invariant measurement that compensates for local geometric differences in the precinct configuration. In other words, the weighted score $\sum_{e\in\C}\w(e)\cdot 1$ is approximately proportional to the number of cut edges there would be if the precincts formed a regular lattice across the state.  For example, if an infinitesimal change in a precinct boundary transformed precincts meeting at a point into adjacent precincts, then the measurement $\sum_{e\in\C}\w(e)\cdot 1$ would credit this change with only an infinitesimal portion of a new cut edge rather than a full cut edge.  Figure~\ref{F:waist} shows the waist of three regular lattices.
\end{enumerate}

In summary, instead of regarding $\tlL$ as a modification of $\LL$ that makes it more similar to $|\C|$, we can regard it as the opposite -- a \emph{weighted cut edge count} that attempts to measure the number of cut edges there would be if the precincts were equipopulous and formed a regular lattice across the state.

We have obtained the following sequence of compactness scores that naturally interpolate between the total perimeter and the cut edge count.
\begin{equation}\label{E:sequence}\LL \stackrel{.75}{\longleftrightarrow} 
\tL\stackrel{.98}{\longleftrightarrow}\tlL\stackrel{.94}{\longleftrightarrow}|\C|.\end{equation}
The numbers in Equation~\ref{E:sequence} are the correlation coefficients between the subsequent scores for a standard ReCom ensemble of 50000 North Carolina congressional maps.  These correlations show how much difference the steps make for a typical application.  For example, $\tL$ and $\tlL$ are very strongly correlated in this application, which means it mattered very little whether the slack was incorporated into the measurement.  We will explain a reason for this observation in the Section 5.

Since $\LL = \sum_{e\in\C}\textcolor{red}{l(e)}$ and $|\C| = \sum_{e\in\C}\textcolor{red}{1}$, our stepping stone scores between them arose by observing the following precise relationship between these two summands:
\begin{equation}\label{E:summand1} \textcolor{red}{l}\cdot \left(\sqrt{\phi}\cdot\frac{1}{\sss}\cdot \frac{1}{\n}\cdot\frac{1}{\w}\right) = \textcolor{red}{1},\end{equation}
and then deciding which of the factors in Equation~\ref{E:summand1} should be incorporated into our stepping stone scores.  Although this decision was somewhat arbitrary -- for example, we could have made the linear change before making the conformal change -- it is natural to keep the factor $\w\cdot\n$ out of our stepping stone scores.  This makes the scores more lattice-independent, which might be considered a desirable property.

\section{Conformal variants of other shape-based compactness scores}
In this section, we consider the effect of the conformal change from the previous section on other shape-based compactness scores.

First, the Polsby-Popper score of a district $D$ is defined as
$$\PP(D) = 4\pi\cdot\frac{\text{area}(D)}{\text{perim}(D)^2}\in(0,1].$$
A common variant of the Polsby-Popper score, called the \emph{Schwartzberg score}, is defined as
\begin{equation}\label{E:schw}\Schw(D) = \PP(D)^{-1/2} = \frac{\text{perim}(D)}{\sqrt{4\pi\cdot\text{area}(D)}}\in[1,\infty).\end{equation}
In order to measure the compactness of the entire map, let $\PP$ and $\Schw$ denote the sum of these scores over the districts of the map.  

We next denote by $\tPP$ and $\tSchw$ the result of measuring these scores after applying the conformal change.  More specifically, each district's conformally-changed area equals its \emph{population}, while its conformally-changed perimeter is computed as in the previous section (with the exception that the discrete approximation of $\phi$ occurs over a single precinct, rather than over a pair of precincts, for segments along the exterior boundary of the state).

\begin{figure}[bht!]\centering
\includegraphics[width=5in]{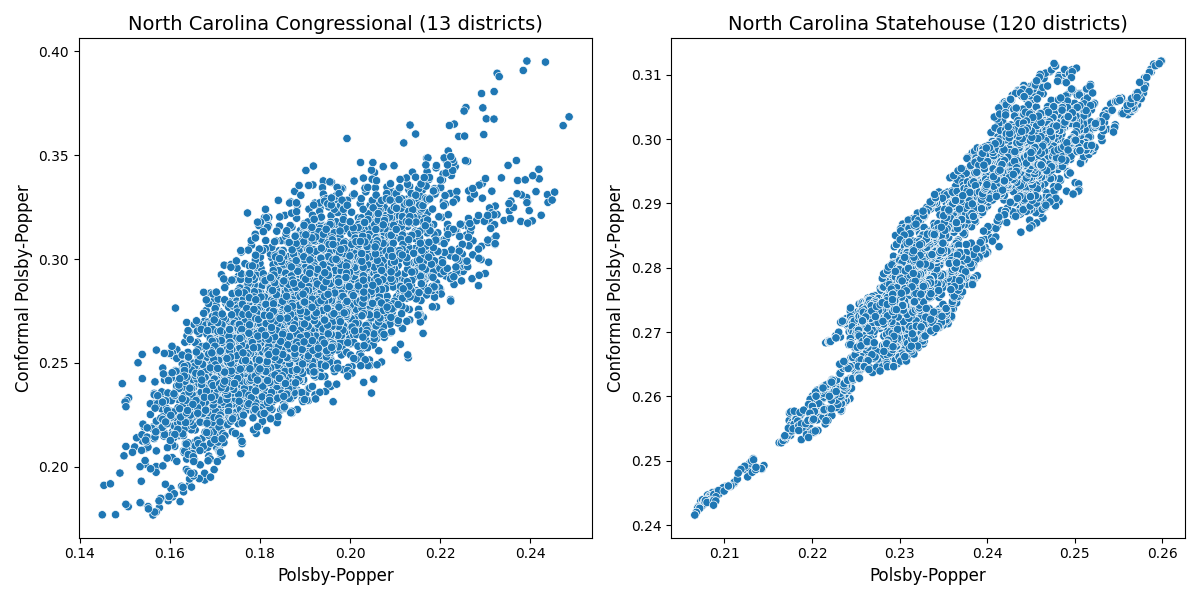}
\caption{$\tPP$ verses $\PP$ for the maps in standard ReCom ensembles.}\label{F:Conf_PP}
\end{figure}

Figure~\ref{F:Conf_PP} suggests that $\PP$ and $\tPP$ are more strongly correlated when $k$ is large (so there is a larger number of smaller districts).  To explain this, recall that the Polsby-Popper score is scale-invariant, so if the population-density function $\phi$ is constant over the precincts of an individual district $D$, then its Polsby-Popper score is unaffected by the conformal change.  This is more nearly true for small NC statehouse districts than for large NC congressional districts. 

The districts are legally required to be approximately equipopulous.  If we assume that they are \emph{exactly} equipopulous, then the conformally modified $\text{area}(D)$ in Equation~\ref{E:schw} equals the population, which is constant over the districts, and hence:
$$\tSchw = c_1\cdot \tL + c_2$$
for constants $c_1$,$c_2$, where $c_2$ involves the perimeter of the boundary of the state.  In summary, after applying a conformal change, a cosmetic variant of the Polsby-Popper score becomes a linear function of the total perimeter score.

The \emph{Reock} score of a district is another fairly common compactness score.  It equals the area of the district divided by the area of its smallest circumscribing circle.  This is an inherently planar definition,  whereas the conformal change turns the state into (a discrete approximation of) a general two-dimensional Riemannian manifold.  Although Riemannian Geometry does provide tools that could be used to define and compute a minimal circumscribing circle, the conformal version of the Reock score might not be geometrically natural enough to warrant the complexity required to compute it.  Similarly, compactness scores based on the convex hull of a district do not seem to have very natural conformal variants.

\section{Modifying ReCom to improve compactness scores}
In this section, we describe variants of ReCom that aim to generate ensembles of maps that score well on shape-based compactness measurement.  For this, we will borrow the tree-weighting strategy that was successfully used in~\cite{Colorado} to generate county-respecting ensembles.

Specifically, suppose that $\lambda$ is a function from the edge set of $\G$ to $[0,\infty)$; that is, it associates every edge $e$ of the state's precinct dual graph with a real number $\lambda(e)\geq 0$.  One key step of the ReCom algorithm is to construct a random spanning tree on the subgraph of $\G$ corresponding to a pair of merged districts.  This is done using Kruskal's algorithm to select a spanning tree of \emph{maximum} weight with respect to uniformly random edge weights; that is, for each edge $e$, its weight $w(e)$ is chosen uniformly from $[0,1]$.  Our modified version of ReCom instead chooses $w(e)$ uniformly from $[0,\lambda(e)]$.  With this modification, the resulting spanning tree is more likely to include higher-$\lambda$ edges, which means that after an edge of the tree is cut, the cut set between the resulting two components is more likely to include lower-$\lambda$ edges.

This strategy was implemented in~\cite{Colorado} using the following choice for the function $\lambda$:
\begin{equation}\label{E:20}\lambda(e) = \ \begin{cases} \text{1,} & \text{if the nodes of $e$ lie in different counties} \\ \text{20,} & \text{otherwise}. \end{cases}
\end{equation}
This results in tree edges that are less likely (and hence cut edges that are more likely) to correspond to county boundaries, and thus district boundaries that are more likely to follow county boundaries.

To target maps that score well on shape-based compactness measurements, we consider the following additional choices for the function $\lambda$:
\begin{enumerate}
\item (Perimeter) $\lambda(e) = l(e)$.
\item (Slack) $\lambda(e) = \sss(e) -1= \frac{l(e)}{d(e)}-1$. 
\end{enumerate}
The reason for subtracting $1$ from the slack is to shift its range from $[1,\infty)$ to $[0,\infty)$, which magnifies the effect.  On the other hand, the effect of any function $\lambda$ could be muted by adding a constant to it, although this paper does not use a muting parameter.

We built an ensemble of $50000$ North Carolina congressional maps ($k=13$ districts) using each of the following functions: Standard ReCom ($\lambda = 1$), county (Equation~\ref{E:20}), perimeter ($\lambda = l$), and slack ($\lambda = \sss-1$).  We then did the same for North Carolina state senate ($k=50$) and North Carolina state house ($k=120$).  We repeated all of this for Florida ($k\in\{12, 40, 120\}$) and several other states. In all cases, the atomic units were VTDs and the population tolerance was $5\%$.

For simplicity of exposition, we will first discuss the North Carolina congressional ensembles in depth, and then consider what is different for other chambers and other states.

\begin{figure}[bht!]\centering
\includegraphics[width=5in]{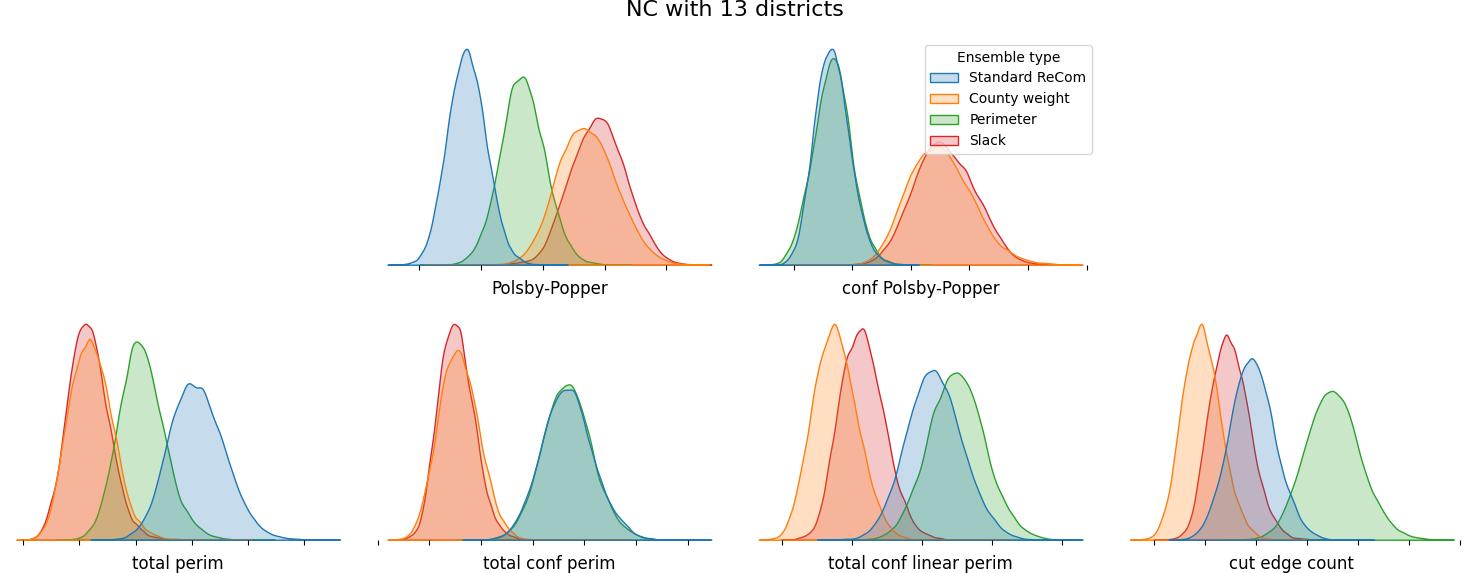}
\caption{The kde plots of several compactness scores for ReCom NC congressional ensembles built with four choices of $\lambda$.  Higher scores are better for  $\PP$ and $\tilde{\PP}$, while lower scores are better for the other measurements.}\label{F:kde}
\end{figure}

Figure~\ref{F:kde} shows how the maps in the NC congressional ensembles scored with respect to several compactness scores: Polsby-Popper ($\PP$), conformal Polsby-Popper ($\tilde{\PP}$), total perimeter ($\LL$), total conformal perimeter ($\tL$),  total conformal linear perimeter ($\tlL$), and cut edge count ($|\C|$).  The bottom row of plots is ordered to match the sequence in Equation~\ref{E:sequence} to highlight how the middle plots interpolate between the plots for $\LL$ and $|\C|$.

\begin{figure}[bht!]\centering
\includegraphics[width=4.5in]{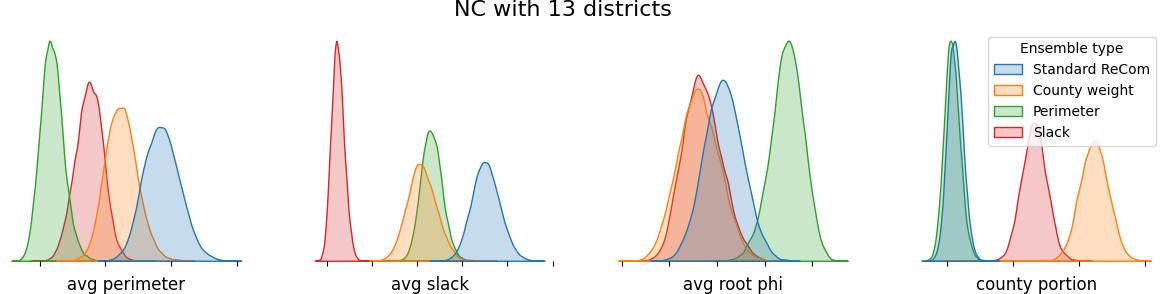}
\caption{The first three graphs are kde plots of the average (over the cut edges) of $l$, $\sss$, and $\sqrt{\phi}$.  The rightmost graph shows the portion of the cut edges that cross county boundaries.}\label{F:NC_three}
\end{figure}

Before discussing these results, one ensemble at a time, we mention that Table~\ref{T:cor} and Figure~\ref{F:NC_three} can together help one interpret and explain the ensembles' compactness scores.  For example, the perimeter ensemble ($\lambda = l$) is designed to be biased in favor of short cut edges (small $l$), which according to Table~\ref{T:cor} are more likely to be straighter (low $\sss$) and contained in more populous regions (high $\phi$).  Figure~\ref{F:NC_three} shows that the perimeter ensemble does indeed turn out this way; compared to the standard ReCom ensemble, the maps of the perimeter ensemble on average have cut edges that are shorter, straighter, and from more populous regions.  In particular, district lines tend to cut through cities.  On the other hand, the rightmost graph in Figure~\ref{F:NC_three} shows that maps in the perimeter ensemble have about the same level of respect for county boundaries as standard ReCom maps, which aligns with the relatively small correlation in Table~\ref{T:cor} between county and $l$.

\begin{table}\caption{Correlation coefficients for several functions on the $7593$ edges of the North Carolina precinct dual graph.  Here `county' equals $0$ for intra-county and $1$ for inter-county edges.}\label{T:cor}
\begin{tabular}{lrrrr}
\toprule
{} &  county &  $l$ &  $\sss$ &   $\phi$ \\
\midrule
county  &      1.00 &       0.05 &  -0.13 & -0.27 \\
$\hspace{.2in}l$ &      0.05 &       1.00 &   0.38 & -0.44 \\
$\hspace{.2in}\sss$     &     -0.13 &       0.38 &   1.00 & -0.11 \\
$\hspace{.2in}\phi$     &     -0.27 &      -0.44 &  -0.11 &  1.00 \\
\bottomrule
\end{tabular}
\end{table}

\begin{enumerate}
\item \textbf{The perimeter ensemble} beat the standard ReCom ensemble with respect to $\PP$ and $\LL$, but lost with respect to $|\C|$.  Biasing towards district boundaries built from small-perimeter segments leads to maps built from a larger number of shorter segments, which on balance seems to improve the total perimeter score. 
\item \textbf{The county ensemble} tied as the best performer with respect to total perimeter.  One reason is that inter-counties edges are straighter (smaller $\sss$) than intra-county edges, as indicated in Figure~\ref{F:NC_three} and Table~\ref{T:cor}.  Moreover, the district lines seem not only to be straighter at the atomic magnification level measured by $\sss$, but at a zoomed-out magnification as well.  By following county lines, the districts seem to partially inherit the nice global shapes of counties, which reduces their cut edge counts and improves their total perimeter scores.

\item\textbf{The slack ensemble} tied with the county ensemble with respect to the total perimeter score and slightly beat it with respect to the Polsby-Popper score.  To achieve this surprising success, it compensating for its slightly larger number of cut edges with a slightly smaller average perimeter per cut edge.  The slack ensemble attempts to build districts out of nearly straight-line segments.  Low-slack edges tend to be shorter (smaller $l$) and are more likely to be county boundaries, as indicated by Figure~\ref{F:NC_three} and Table~\ref{T:cor}.  Although these maps do not follow county boundaries as closely as maps in the county ensemble, they seem to follow closely enough to inherit some of the benefits.  On the other hand, both the slack and county ensemble maps are forced to split some counties, and the maps from the slack ensemble seems to benefit from doing so along straighter paths.
\end{enumerate}

We tried two additional choices for the function $\lambda$ that ended up performing too poorly to merit inclusion in our figures.  One of these under-performing choices was $\lambda = \sqrt{\phi}\cdot l$ (the conformal perimeter).  This ensemble performed worse than standard ReCom with respect to $|\C|$ and a bit better with respect to the other scores.  However, it performed worse than both the county and the slack ensembles with respect to all of the scores, including the total conformal perimeter score, which it seems custom-built to improve.  The other under-performing choice we tried was $\lambda = \sqrt{\phi}\cdot d$ (the conformal linear perimeter), which similarly lost to the slack and county ensembles, even with respect to the total conformal linear perimeter score, which it seems designed to improve.  These two under-performing ensembles were very similar to each other.

\begin{figure}[bht!]\centering
\includegraphics[width=2.5in]{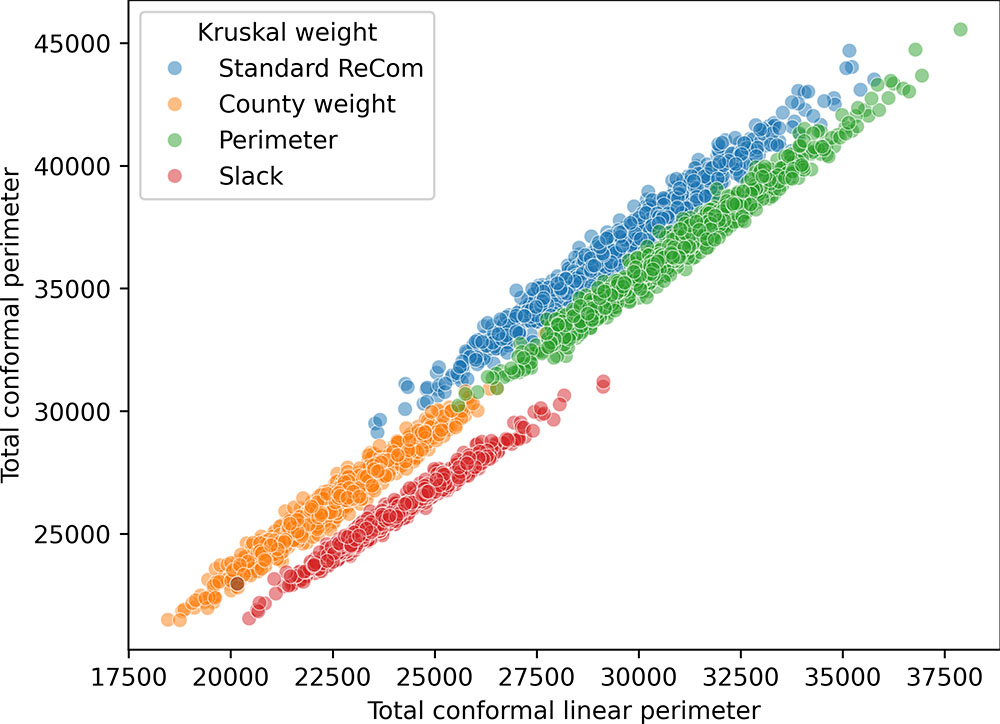}
\caption{Scatter plot of $\tL$ verses $\tlL$ for the maps of NC congressional ensembles.}\label{F:scatter}
\end{figure}

Before moving to other chambers and other states, we pause to look at the strong correlation of $\tL$ and $\tlL$ that we previously encountered in Equation~\ref{E:sequence} with respect to the standard ReCom ensemble.  Figure~\ref{F:scatter} shows a strong correlation with respect to each ensemble individually.  The different regression line slopes for the four different ensembles reflect their differences with respect to the average slack (Figure~\ref{F:NC_three}).

\begin{figure}[bht!]\centering
\includegraphics[width=5in]{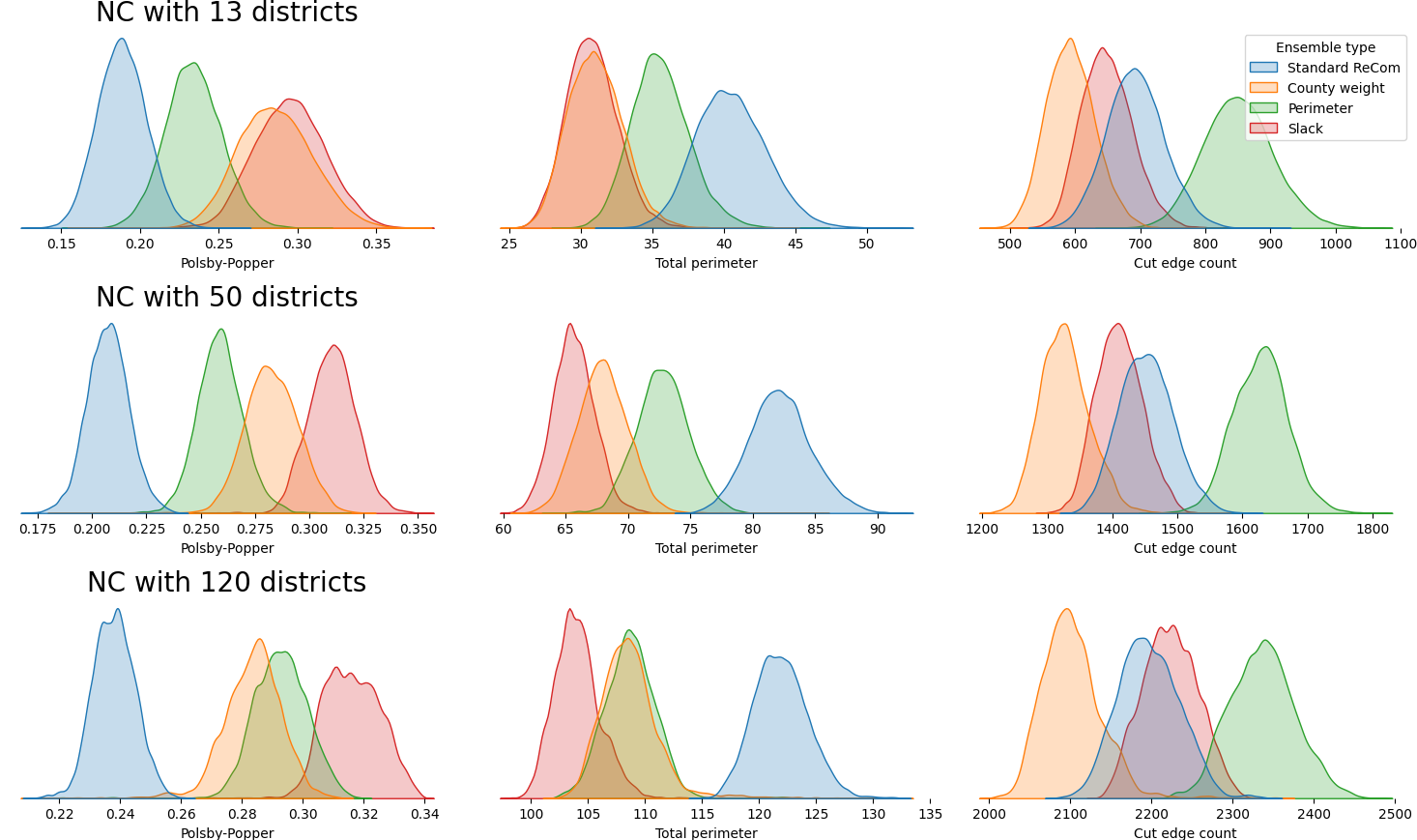}
\caption{The kde plots of three compactness scores for NC congressional (top), senate (middle), and statehouse (bottom) ensembles.  The Polsby-Popper score is averaged (rather than summed) over the districts.}\label{F:NC_3_compact}
\end{figure}

\begin{figure}[bht!]\centering
\includegraphics[width=5in]{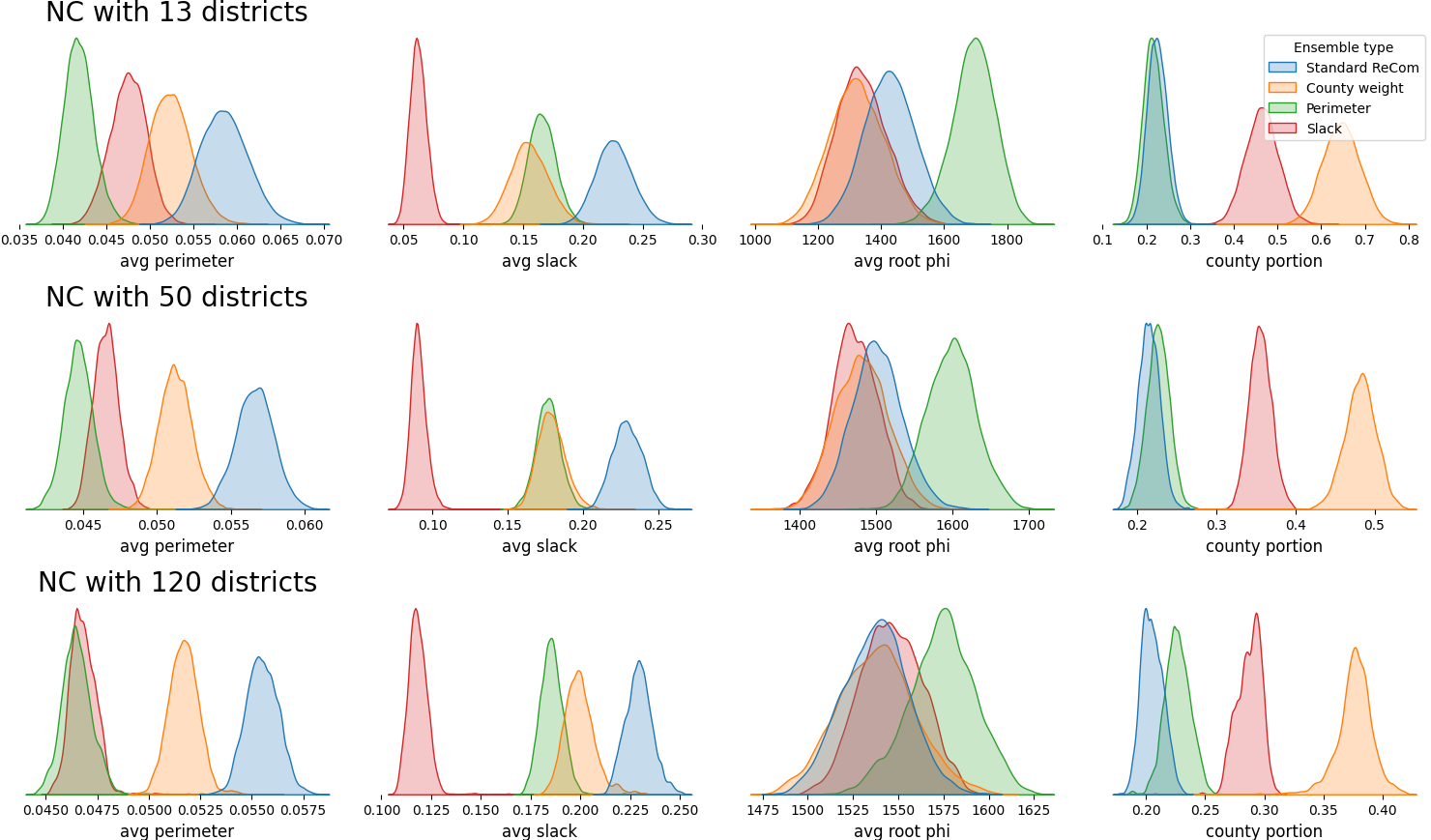}
\caption{Expansion of Figure~\ref{F:NC_three} to also include senate and statehouse ensembles.}\label{F:NC_3_avg}
\end{figure}

We end this section by considering how our methods perform in other chambers and other states. Figure~\ref{F:NC_3_compact} displays the three most common of the six compactness scores from Figure~\ref{F:kde} (left = $\PP$, middle = $\tL$, right = $|\C|$), not only for NC congressional (top), but also for NC senate (middle) and NC statehouse (bottom).  Figure~\ref{F:NC_3_avg} similarly expands Figure~\ref{F:NC_three} to include these other legislative chambers.

The takeaway is that increasing the number of districts improves the performance of the slack and perimeter ensembles (relative to the standard ReCom), but worsens the performance of the county ensemble.  In particular, with respect to the Polsby-Popper score, the slack and county ensembles tie for first place when $k=13$.  But when $k=120$, the slack ensemble pulls ahead to first place, and the perimeter ensemble catches up with the county ensemble.  The total perimeter score tells the same story.

One explanation for the worsening performance of the county ensemble as $k$ increases is the limited number of available inter-county edges.  The bottom-right plot of Figure~\ref{F:NC_3_compact} shows that \emph{all} statehouse maps from \emph{all} ensembles have at least 2000 cut edges, whereas the NC dual graph only has 1303 inter-county edges.  Moreover, the maps aren't able to use all of these inter-county edges because of population and geographic constraints; in particular, the maps in the statehouse county ensemble use $55\%-65\%$ of them. 

We additionally created ensembles for other states including Ohio, Pennsylvania, and Florida.  The resulting plots were too similar to North Carolina's plots to merit inclusion in the paper, but are available in the GitHub repository.  The only difference of note was that the correlation coefficient between county and $\sss$ from Table~\ref{T:cor} was much stronger in NC than in the other states (NC = $-.12$, OH = $-.04$, PA = $-.02$, FL = $-.01$).  Nevertheless, all of the plots looked about the same for the other states as for NC, including the ``county portion'' plots in Figure~\ref{F:NC_3_avg} (right).  Apparently, the maps of the slack ensembles manage to include a lot of county boundary segments, even though such segments are not much straighter than their intra-county counterparts. 

\section{Conclusion}
It is very natural to calculate compactness scores of a map after applying a conformal change that scales the underlying metric of the state by population density.  Among other things, this strategy helps quantify the differences between the shape-based total perimeter score $\LL = \sum_{e\in\C}\textcolor{red}{l(e)}$ and the discrete cut edge count $|\C| = \sum_{e\in\C}\textcolor{red}{1}$.  More specifically, since
\begin{equation}\label{E:summand} \textcolor{red}{l}\cdot \left(\sqrt{\phi}\cdot\frac{1}{\sss}\cdot \frac{1}{\n}\cdot\frac{1}{\w}\right) = \textcolor{red}{1},\end{equation}
these scores differ with respect to $\phi$, $\sss$, $\n$ and $\w$.

We used this observation to build intermediate scores interpolating between $\LL$ and $|\C|$.  The score $\tL$ incorporated $\phi$, while the score $\tlL$ also incorporated $\sss$, and thus could be thought of as a weighted version of $|\C|$ that attempts to remove its dependence on the precinct tiling ($\n$ and $\w$).

The second idea of this paper is to achieve better shape-based compactness scores by nudging the ReCom algorithm towards maps built from \emph{shorter} segments or \emph{straighter} segments.  The resulting ``perimeter'' and ``sack'' ensembles performed better than standard ReCom, especially when the number of districts is large.  In fact, when the number of districts is large, the slack ensemble also outperformed the county-respecting variant of ReCom from~\cite{Colorado}.  It is interesting that, even without any county awareness, there are natural modifications of ReCom that build ensembles of maps with significantly better compactness scores.
\bibliographystyle{amsplain}

\end{document}